\documentclass[12pt]{article}
\usepackage{amsthm,amsmath, graphicx, booktabs, multirow, xcolor, amssymb, tikz, adjustbox, epstopdf, upgreek}
\usepackage{latexsym}
\usepackage{graphicx}
\usepackage{enumitem}
\usepackage{fullpage}
\usepackage{setspace}
\usepackage{longtable}
\RequirePackage{natbib}
\bibpunct{(}{)}{;}{a}{}{,}
\usepackage{dcolumn}
\newcolumntype{.}{D{.}{.}{-1}}
\newcolumntype{d}[1]{D{.}{.}{#1}}
\usepackage[top=.8in, left=.5in, right=.5in,
bottom=.8in]{geometry}
\usepackage[compact]{titlesec}
\usepackage{booktabs}
\usepackage{url} 

\usepackage{color}
\RequirePackage[colorlinks,citecolor=blue,urlcolor=blue]{hyperref}

\newtheorem{theorem}{Theorem}

\usepackage{caption}
\usepackage{subcaption}
\usepackage[linesnumbered, ruled, vlined]{algorithm2e}

\expandafter\def\expandafter\normalsize\expandafter{\normalsize\setlength\abovedisplayskip{0pt}}
\expandafter\def\expandafter\normalsize\expandafter{\normalsize\setlength\belowdisplayskip{0pt}}
\expandafter\def\expandafter\normalsize\expandafter{\normalsize\setlength\abovedisplayshortskip{0pt}}
\expandafter\def\expandafter\normalsize\expandafter{\normalsize\setlength\abovedisplayshortskip{0pt}}

\begin{document}
\pagestyle{plain}

\def\spacingset#1{\renewcommand{\baselinestretch}%
{#1}\small\normalsize} \spacingset{1}


  \title{Relative Contrast Estimation and Inference for Treatment Recommendation }
  \author{Muxuan Liang \thanks{
  	Public Health Sciences Division, Fred Hutchinson Cancer Research Center}\\
  	\and
  	Menggang Yu\thanks{
  	Department of Biostatistics and Medical Informatics, University of Wisconsin-Madison}}
  \maketitle

\thispagestyle{empty}
\abstract{
When there are resource constraints, it is important to rank or estimate treatment benefits according to patient characteristics. This facilitates prioritization of assigning different treatments. Most existing literature on individualized treatment rules targets absolute conditional treatment effect differences as the metric for benefits. However, there can be settings where relative differences may better represent such benefits. In this paper, we consider modeling such relative differences that form scale-invariant contrasts between conditional treatment effects. We show that all scale-invariant contrasts are monotonic transformations of each other. Therefore we posit a single index model for a particular relative contrast. Identifiability of the model is enforced via an intuitive $l_2$ norm constraint on index parameters. We then derive estimating equations and efficient scores via semiparametric efficiency theory. Based on the efficient score and its variant, we propose a two-step approach that consists of minimizing a doubly robust loss function and a subsequent one-step efficiency augmentation procedure to achieve efficiency bound. Careful theoretical and numerical studies are provided to show the superiority of the proposed approach. 
}

\newcommand{\n}{\noindent}

\newpage
\section{Introduction}
\label{sec:intro}

In this era of precision medicine,  methods for constructing individualized treatment rules are flourishing \citep{Zhao2012, Xu2015, Chen2017, zhao2019, Liang2020}. An individualized treatment rule is a decision rule that maps from the space of the covariates $\mathcal{X}$ to the interventions
 $\mathcal{T}=\left\{-1, 1\right\}$. Given a decision rule $d$ and patient covariates $X$, treatment recommendation is given by the sign of $d(X)$. 
Let $(Y_1, Y_{-1})$ be the potential outcomes corresponding to the treatments $T=1$ and $T=-1$, respectively \citep{rubin1974, rubin2005}. The individualized treatment effect is commonly reflected through an absolute difference defined as $\Delta(X)=\mu_1(X)-\mu_{-1}(X)$ where $\mu_1(X) = E(Y_{1}\mid X)$ and $\mu_{-1}(X) = E(Y_{-1}\mid X)$. The optimal individualized treatment rule that aims to select the best treatment for each individual is therefore $d_{\rm{opt}}(X)={\rm sign}\{\Delta({X})\}$.

Due to this relationship, many methods for estimating the optimal individualized treatment rule focus on estimation of $\Delta(X)$.  Q-learning is an intuitive approach that estimates $\Delta(X)$ by positing models for  both $\mu_1(X)$ and $\mu_{-1}(X)$ \citep{Watkins1992, chakraborty2010inference, qian2011, Laber2014}. A-learning on the other hand directly models $\Delta(X)$ \citep{murphy2003, lu2013, shi2016, shi2018, song2017} therefore typically makes fewer modeling assumptions.  In particular, \cite{song2017} proposed a single index model for $\Delta(X)$. \cite{Liang2020} considered a general semiparametric modeling approach for $\Delta(X)$. However, both approaches are not semiparametrically efficient for parameter estimation due to the difficulty of directly solving the efficient score. Although the method of \cite{Liang2020} is locally efficient, the coefficients need to satisfy a Grassmann-manifold condition for model identifiability. The condition requires a prior knowledge of at least one non-zero coefficients. On the other hand, \cite{song2017} adopted a more natural $l_2$-norm constraint which avoids such requirement. However their theoretical development overlooked this $l_2$-norm constraint and led to an overly optimistic asymptotic variance. When the $l_2$-norm constraint is imposed, the asymptotic variance is de-generated, thus, creating challenges for inference.

When resource constraints prohibit assigning all patients to their optimal treatments, it becomes necessary to prioritize assignment according to the  magnitudes or ranks of the individualized treatment effects, e.g. from an estimated  $\widehat{\Delta}(X)$. For example, if $\widehat{\Delta}(X_1)> \widehat{\Delta}(X_2)$, then the patient with ${X_1}$ has high priority than the patient with ${X_2}$. Alternatively, one can also use classification based approaches and recommend subjects only when their estimated treatment benefits surpass some threshold \citep{Ming2016}. 

However, there can be many settings where relative differences are better individualized metrics to reflect treatment benefits for outcomes such as cost saving, quality of life improvement, and disease response rate. For example, a fixed amount of saving can carry different levels of importance for subjects with different levels of wealth. An improvement of 5\% response rate also can be conceived differently depending on the baseline response rate. Decrease of a fixed amount of health measurements such as blood pressure, blood sugar level, and cholestol level is also most meaningful when baseline measurements are considered. Consequently, \citet{Yadlowsky2020} proposed an exponential regression model for a ratio-based contrast in analyzing a multiple sclerosis data set.
 
Therefore in this paper, we study relative contrasts on non-negative outcomes whose formal definition is in Section \ref{sec:gen_contrast}. Typical examples of relative contrasts include the following ratio-based contrast 
\begin{eqnarray}
\theta(X) = \log \{\mu_{1}(X)/{\mu_{-1}(X)}\} \label{ratio-RC}
\end{eqnarray} {that is used by \citet{Yadlowsky2020}}, and weighted contrasts $\Delta(X)/\mu_{-1}(X)$ and $\Delta(X)/S(X)$ where $S(X) = \mu_1(X)+\mu_{-1}(X)$.
In Theorem \ref{thm1} of Section \ref{sec:gen_contrast}, we will show that all relative contrasts are linked in the sense that they are all monotonic transformations of each other. For example, ${\Delta(X)}/{\mu_{-1}(X)} =  e^{\theta(X)}-1$ and $\Delta(X)/S(X)=\left\{e^{\theta(X)} -1\right\}/\left\{e^{\theta(X)}+1\right\}$.

 Relative contrasts have additional interesting properties. First, they have the same signs as $\Delta(X)$ when the potential outcomes are positive. As such, the optimal individualized treatment rule can also be written as $d_{\rm{opt}}(X)={\rm sign}\left\{\Delta(X)\right\}={\rm sign}\left\{\theta(X)\right\}$, for example. Second, relative contrasts typically reflect the treatment effect or contrast $\Delta(X)$ weighted by baseline outcome levels such as $\mu_{-1}(X)$ and $S(X)$. A patient with a low level of $Y_{-1}$ may have a low level of $Y_1$. As such, the contrast function would be small but the percentage of changes may not be small.

Another noticeable fact about the relative contrasts is that they are closely connected with existing works. \citet{Zhao2012} pioneered an outcome weighted learning (O-learning) approach for individualized treatment rule from a classification perspective.
The work spurred many subsequent papers on extending this method \citep{Zhao2012, Zhang2012b, Tian2014,Xu2015, Chen2017, zhao2019}. In particular, this type of approach aims to maximize the following value function which is the expected value of the potential outcome that uses an decision rule $d(X)$, that is,
$E\{Y_{d(X)}\}=E[\pi_T^{-1}(X)Y I\{T=d(X)\}]$,
	where $\pi_T(X)= pr(T\mid X)$ is known as the propensity score \citep{Rosenbaum1983}.
	\citet{Ming2016} extended this classification to incorporate a reject option and pointed out that the O-learning type of approaches are basically trying to determine subsets determined by ${\Delta(X)}/S(X)$. In other words, ${\Delta(X)}/S(X)$ is implicitly involved in the solutions of O-learning type of approaches.


We therefore explicitly model relative contrasts in this article. In particular, we propose single index models $\psi_0(X^\top\beta_0)$ for the relative contrasts where $\psi_0(\cdot)$ is an unknown monotone function and $\beta_0$ is a $p\times 1$ coefficient vector. This modeling choice draws from the facts that all relative contrasts are monotonic transformations of each other and that monotonic transformations of single index models are still single index models. In order to recover the rank of relative contrasts, we just need to infer $\beta_0$ when $\psi_0$ is monotonic.

Similar to \cite{song2017}, we use an $l_2$-norm constraint on the index parameter vector $\beta_0$ for identifiability. However we give proper attention to this constraint in our theoretical development. We also directly work with the semiparametrically efficient score for estimating $\beta_0$. To overcome the computational difficulties associated with  directly solving the efficient score due to multiple nuisance parameters, we adopt a two-step approach. In the first step, motivated by a simplified efficient score, we propose a doubly robust loss to estimate relative contrasts. In the second step, we improve the estimated coefficient vector in the first step by a one-step efficiency augmentation step. The improved estimator is shown to obtain the semiparametric efficiency lower bound. 
\section{Modeling the relative contrasts}
\label{sec:gen_contrast}

We define a {\em relative contrast function} (RCF)  $h:\mathbb{R}^+\times \mathbb{R}^+\to \mathbb{R}$ as a continuous function on positive numbers satisfying 
\begin{enumerate}[label=(\alph*), leftmargin=.75in]
	\item \label{condition:a} $h(a, a)=0$;
	\item \label{condition:b} $h(a, b)$ is strictly increasing in $a$ for any fixed $b$; 
	\item \label{condition:c} $h(\lambda a, \lambda b)=h(a, b)$, for all $\lambda >0$. 
\end{enumerate} Corresponding to a given relative contrast function $h$,  $h\left (\mu_1(X) , \mu_{-1}(X) \right )$ is a  relative contrast.  

Property~\ref{condition:a} imposes that if $\mu_1(X) =\mu_{-1}(X)$, we have $h\left(\mu_1(X) , \mu_{-1}(X) \right)=0$.  When $h$ is differentiable in its first argument, Property~\ref{condition:b} is equivalent to ${\partial h(a, b)}/{\partial a}>0$ for a fixed $b$. This property also guarantees that $\Delta(X)>0$ is equivalent to $$h\left(\mu_1(X) , \mu_{-1}(X) \right)>0.$$ Property~\ref{condition:c} implies that $h$ is invariant to scale changes of the outcomes. 

Many current methods comparing $\mu_1(X)$ and $\mu_{-1}(X)$ do not satisfy Property~\ref{condition:c}. The absolute difference $\Delta(X)$ corresponds to $h(a, b) = a-b$ which fails to satisfy Property~\ref{condition:c}. In \citet{wenchuan2018}, the authors proposed a new casual estimand for binary response called covariate-specific treatment effect, defined as
$
\mathrm{logit}\{\mu_{1}( X)\}-\mathrm{logit}\{\mu_{-1}( X)\},
$
where $\mathrm{logit}(u)=\log u - \log(1-u)$. The covariate-specific treatment effect corresponds to the choice of $h(a,b)=\mathrm{logit}(a)-\mathrm{logit}(b)$ and it does not satisfy Property~\ref{condition:c}.

On the other hand, one can have many possible choices of relative contrast functions. For example, both $h(a, b)=\log (a/b)$,  $h(a,b)=(a-b)/b$ and $h(a,b)=(a-b)/(a+b)$ are relative contrast functions. Their corresponding relative contrasts are 
$\theta(X)$ defined in \eqref{ratio-RC}, ${\Delta(X)}/{\mu_{-1}(X)}$ and ${\Delta(X)}/S(X)$ where $S(X) \equiv \mu_1(X)+\mu_{-1}(X)$. In practice, the choice of $h$ can vary upon the scientific needs. 


Despite many possible choices of relative contrast functions, the following theorem indicates that they are all intrinsically related in the sense that their corresponding relative contrasts are monotonic transformations of each other. 
\begin{theorem}\label{thm1}
	For any two relative contrast functions $h$ and $\widetilde{h}$,  there exists a strictly increasing mapping $g$ such that $\widetilde{h}\left (\mu_1(X), \mu_{-1}(X)\right )=g\{{h}\left (\mu_1(X), \mu_{-1}(X)\right )\}$.
\end{theorem}

Theorem~\ref{thm1} is proved in the online supplemental materials. It motivates us to model relative contrasts using single index models. Suppose that there exists an relative contrast based on $h$ that follows a single index model $h(\mu_1(X), \mu_{-1}(X))=\psi(X^\top\beta)$, where $\psi$ is an increasing function and $\beta$ is a $p$-dimensional coefficient vector. From Theorem~\ref{thm1}, for any other relative contrast function $\widetilde{h}$, there exists a strictly increasing $g$ such that $\widetilde{h}(\mu_1(X), \mu_{-1}(X))=g\circ \psi(X^\top\beta)$. As such, a single index model assumption for one relative contrast is equivalent to single index models for all relative contrasts.

Therefore we focus on modeling just one relative contrast,  $\theta(X)=\log\{\mu_{1}(X)/\mu_{-1}(X)\}$ corresponding to $h(a, b)=\log(a/b)$. We choose to model $\theta(X)$ because its range is $(-\infty, \infty)$ so that we do not need to put range constraints on the wrapper function of the single index model.   In the rest of the article, we will focus on the following model: 
\begin{eqnarray}
\theta(X)=\log \bigg \{\mu_{1}(X)/\mu_{-1}(X) \bigg \}=\psi_0(X^\top\beta_0), \label{the-model}
\end{eqnarray} 
where $\psi_0$ is an increasing unknown wrapper function and $\beta_0$ is a $p$-dimensional vector.  Note that because $\psi_0$ is unknown,  the single index $\beta_0$ is not identifiable up to a positive multiplicative scale. Therefore we force $\|\beta_0\|_2=1$ to make it identifiable similar to  \citet{Carroll1997backfitting, Liang2010profile, Ma2016}, and \citet{song2017}.
\section{Estimation}
\label{sec:method_db}

In this section, we first derive the space of all influence functions for estimation of $\theta( X)$ and propose a doubly robust estimating equation for $ \beta_0$  and $\psi_0$.
To possible improve efficiency from the doubly robust estimating equation, we further propose a semiparametrically efficient estimating equation for ${  \beta_0}$.  

We assume the usual causal conditions on the potential outcomes and treatment assignment mechanism \citep{Rosenbaum1983, Imbens2015} to facilitate likelihood derivation. Specifically, we assume 1) the Stable Unit Treatment Value Assumption (SUTVA), that is, the potential outcomes of any patient are not affected by the treatment assignments of other patients; 2) observed outcome consistency: $Y=Y_T$, that is, the observed outcome is the corresponding potential outcome;  and 3) no unmeasured confounding: $(Y_{-1}, Y_1)\perp T|  X$, that is, all dependences between the potential outcomes and treatment assignment are captured by $  X$. 

\subsection{Orthogonal tangent space and efficient score}
\label{sec: efficiency_theory}

We start from derive the orthogonal tangent space and efficient score for estimating $\beta_0$, which motivate our two-step procedure. We assume the usual causal conditions on the potential outcomes and treatment assignment mechanism \cite{Rosenbaum1983, Imbens2015} to facilitate likelihood derivation. Specifically, we assume 1) the Stable Unit Treatment Value Assumption, that is, the potential outcomes of any patient are not affected by the treatment assignments of other patients; 2) observed outcome consistency $Y=Y_T$, that is, the observed outcome is the corresponding potential outcome;  and 3) no unmeasured confounding $(Y_{-1}, Y_1)\perp T\mid X$, that is, all dependences between the potential outcome and treatment assignment are captured by $X$. The no unmeasured confounding assumption is also known as strong ignorability.

Under the model assumption~\eqref{the-model}, the following theorem characterizes the orthogonal tangent space of estimating $\beta_0$.

\begin{theorem}\label{thm:influence_single_index}
	Under Model \eqref{the-model}, the orthogonal tangent space of $\pi_T( X)= pr(T\mid X)$, $S= \mu_1(X)+\mu_{-1}(X)$, and $\psi_0$ is
	\begin{equation*}
	\left\{\pi_T^{-1}T\big \{Y(1+e^{-T\psi_0(X^\top\beta)})-S\big \}\left\{\alpha(X)-{E}\left(S\alpha(X)\mid X^\top\beta \right)/{E}\left(S\mid X^\top\beta\right)\right\}: \alpha(X)\in L_2(X)\right\}.
	\end{equation*}
\end{theorem}

\medskip

By projecting the score function onto the orthogonal tangent space, we can obtain the efficient score for $\beta_0$ as follows.
\begin{theorem}\label{thm:efficient_score}
Denote $\epsilon \equiv Y-  {E}(Y\mid {X}, T)$ and $\sigma_T^2(X)=E(\epsilon^2\,\mid\,T, X)$. The efficient score for $\beta$ under Model \eqref{the-model} is
\begin{equation*}
\pi_T^{-1}T\big \{Y(1+e^{-T\psi_0(X^\top\beta)})-S\big \}\nabla\psi_0\, \alpha_*(X),
\end{equation*}
where
\begin{eqnarray*}
\alpha_*(X)&=&V(X)S\left\{X-{E}\left(V(X)S^2X\mid X^\top\beta\right)/{E}\left(V(X)S^2\mid   X^\top\beta\right)\right\},
\end{eqnarray*}
and
\begin{eqnarray*}
V^{-1}(X)&=&\left\{\pi_1^{-1}e^{-\psi_0(X^\top\beta)}\sigma_1^2(X)\,+\pi_{-1}^{-1}e^{\psi_0(X^\top\beta)}\sigma_{-1}^2(X)\right\}\left\{e^{\psi_0(X^\top\beta)/2}+e^{-\psi_0(X^\top\beta)/2}\right\}^{2}.
\end{eqnarray*}
\end{theorem}

Therefore, the key element in using the efficient score requires evaluation or estimation of $\pi_T$, $S$, $\psi_0$,and $\alpha_*(X)$. These nuisance parameters create great difficulties if we aim to solve the efficient score directly. First, the $\alpha_*(X)$ depends on $V(X)$, ${E}(V(X)S^2\mid  X^\top\beta)$, and $ {E}(V(X)S^2X\mid X^\top\beta)$. More importantly, the $\alpha_*(X)$ involves $\beta$ which is of primary interest. Second, the gradient of the efficient score is not of full rank due the constraint $\|\beta\|_2=1$. Thus, we consider an alternative solution to avoid direct solving the efficient score - a two-step approach. In the first step, we estimate an initial value of $\beta_0$ through a doubly robust loss function. In the second step, we use the efficient score to form a one-step efficiency augmentation estimation.

\subsection{The first step: Minimize a doubly robust loss function}
\label{sec:db_robust}

In this section, we propose a loss function to estimate $\theta(X)$ (or $\beta_0$) with double robustness -  the minimizer of the proposed loss is $\theta(X)$ if either propensity sore or outcome regression model is correct. In order to construct such loss function, we start with a simplified efficient score. More specifically, we misspecify $\alpha_*(X)$ as $X$ and consider the following estimating equation,
$
	\pi_T^{-1}T\big \{Y(1+e^{-T\psi_0( X^\top\beta)})-S\big \}\nabla\psi_0  X.
$
Other misspecification for the efficient score can also be considered. For example, the estimating equation of \citet{Yadlowsky2020} corresponds to the choice of $$\alpha(X)=X \pi_1\pi_{-1}\left\{1+e^{\theta(X)}\pi_1+e^{-\theta(X)}\pi_{-1}\right\}^{-1}.$$ Our modification to the efficient score shares two benefits. First, the parameter $\beta$ concentrates in only one part of the estimating equation and does not involve in $\pi_T$, $S$, or $\psi_0$. Second, it can be solved by minimizing a loss function as introduced in following theorem. We have the following theorems whose proofs can be found in the online supplemental materials.

\begin{theorem}\label{thm:loss_function}
	Either $\pi_T$ or $S$ is correct, under the assumption that $ {E}(Y\mid X)>0$, $\theta(X)$ is the unique minimizer (almost surely) of the loss function
	\begin{equation}\label{eq:proposed_loss}
		 {E}(\ell(f(X); {\pi}_T, {S}))
	\end{equation}
	over all bounded measurable functions, where  
\begin{equation*}
\ell(f( X) ;{\pi}_T, {S})=\pi_T^{-1}{Y}\exp\{-Tf(X)\}+\pi_T^{-1}T(S-Y)f(X).
\end{equation*}
\end{theorem}

Theorem~\ref{thm:loss_function} holds without assuming any structure on $\theta(X)$. Thus, the proposed loss function can be used to estimate $\theta(X)$ under any structure assumption. As a special case, when we assume that $\theta(X)=\psi_0(X^\top\beta_0)$ and $\psi_0$ is known, the first order condition of proposed loss is the simplified efficient score.

Next, we discuss how to use the proposed loss function to estimate $\beta_0$ when $\psi_0$ is unknown. When $\psi_0$ is unknown, to estimate the wrapper function $\psi_0$ in Model \eqref{the-model}, similar to \cite{song2017}, we use monotone B-splines. That is, we let
\begin{equation*}
\psi(t)\approx \sum_{j=1}^{K_n+M} \xi_jN_j(t), \xi_1\leq\cdots,\leq \xi_{K_n+M},
\end{equation*}
where $N_j$'s are B-spline bases, $K_n$ is the number of  interior knots in an bounded interval containing $ X^\top\beta$ and $M$ is the order of the B-spline bases.
%
%
Let $\xi = (\xi_1,\cdots, \xi_{d})^\top$ with $d=K_n+M$. In general observational studies, in addition to ${  \xi}$ and $\beta$, the loss function $\ell(f ;{\pi}_T, {S})$ 
involves two more unknown nuissance parts: $\pi_T$ and $S$. For ease of presentation, we first assume these nuisance parts are known and then consider the setting when they need to be estimated. 

{To ensure the monotonicity of $\psi$, we impose two constraints on $  \xi$ as in \citet{leitenstorfer2006}.  Specifically, let $  B$ be a $(d-1)\times d$ matrix with  $B(i,i)=1$, $B(i,i+1)=-1$, and 0 for other entries. Then the first constraint is $  B \xi \leq 0$ which requires $  \xi$ to be monotonically increasing. The second constraint is on the L-1 norm of $  \xi$: $\sum_{j=1}^{d}|\xi_j|\leq C$ where $C$ is a sufficient large constant.} Combining this approximation with the proposed loss function in ~\eqref{eq:proposed_loss}, we estimate $(\psi_0,\beta_0)$ by
\begin{eqnarray*}
\min_{ \xi,\beta}&&  {E}_n\left(\ell( \xi,\beta;{\pi}_T, S)\right)\\
\textrm{subject to}&&   B \xi \leq 0 \text{, } \sum_{j=1}^{d}|\xi_j|\leq C \text{, and } \|\beta\|_2=1,
\end{eqnarray*}
where
\begin{equation}
\ell( \xi,\beta;\pi_T, S)=\pi_T^{-1}Y\exp\left\{-T\sum_{j=1}^{d} \xi_jN_j( X^\top\beta)\right \}+\pi_T^{-1}T({S}-Y)\sum_{j=1}^{d} \xi_jN_j( X^\top\beta). \label{loss_function2}
\end{equation}

To  solve this optimization problem, we iterate between two steps until convergence. {The iteration stops when the maximum coordinate-wise differences of both $  \xi$ and $\beta$ between two iterations become smaller than a pre-specified constant such as $10^{-3}$}. In the first step, we fix $\beta$ and optimize  over $ \xi$. {The number of knots $K_n$ is chosen based on Theorems~\ref{thm:consistency} and~\ref{thm:asymptotic_normal}. Given the number of knots $K_n$, we choose equally spaced quantiles of  $ X^\top\widehat{\beta}^{(l)}$ to form basis functions in the $l$th iteration.} This step solves a convex optimization problem with linear constraints. In the second step, we fix $ \xi$ and optimize $\beta$ under the constraint $\|\beta\|_2=1$. This step solves a nonlinear optimization problem with a quadratic constraint. The details are included in Algorithm \ref{algorithm:1}. 
\medskip
\begin{minipage}{0.95\linewidth} 
\begin{algorithm}[H]\label{algorithm:1}
	\caption{Estimation of $\psi(  X^\top \beta)$ with {\em known} ${\pi}_T$ and ${S}$.}
	\SetAlgoLined
	\label{algo1:1} Obtain an initial estimator $\hat{ \beta}^{(0)}$: optimize a linear form $\psi(  X^\top \beta)=  X^\top \beta$ and then normalize it with $\|\hat{ \beta}^{(0)}\|_2=1$. Set $l=0$\;
	\label{algo1:2} Given $\hat{ \beta}^{(l)}$, solve the following to obtain $\hat{ \xi}^{(l)}$:
	\begin{eqnarray*}
		\min_{ \xi}  \mathbb{E}_n\left[\ell( \xi,\hat{ \beta}^{(l)}\!\!;\, {\pi}_T, S)\right]
		\quad \textrm{subject to} \quad    B \xi \leq 0\ \text{ and } \sum_{j=1}^{d}|\xi_j|\leq C;
	\end{eqnarray*}
\\
   \label{algo1:3}Given $\hat{ \xi}^{(l)}$, solve the following to obtain $\hat{ \beta}^{(l+1)}$:
    \begin{equation*}
    	\min_{\| \beta\|_2=1} \mathbb{E}_n\left[\ell(\hat{ \xi}^{(l)}\!\!\!,\, \beta;{\pi}_T, S)\right];
    \end{equation*}
 \\
     \label{algo1:4}Iterate Steps 2 and 3 until convergence. Let $\hat{ \xi}$ and $\hat{ \beta}$ be the estimates at the last iteration. We set $\hat{\psi}(\cdot)=\sum_{j=1}^{d} \hat{\xi}_jN_j(\cdot)$\;
\end{algorithm}
\end{minipage}

\bigskip

Now we consider the need to estimate the nuisance parts $\pi_T$ and $S$. Both parametric and non-parametric models can be used for such estimation. For example, parametric logistic regression or nonparametric kernel regression  can be used to estimate $\pi_T$;  Likewise, to estimate $S$, we can estimate the conditional means of outcome $Y$ on $ X$ on the treatment, $\widehat{\mu}_1( X)$, and control groups, $\widehat{\mu}_{-1}( X)$, separately. 
An estimator of $S$ can be obtained by $\widehat{\mu}_1( X)+\widehat{\mu}_{-1}( X)$.

The non-parametric models avoid model mis-specification but usually result in convergence rates slower than $O_p(n^{-1/2})$. To deal with the slow convergence rates, we adopt a cross-fitting procedure similar to \citet{victor2018}. In Algorithm~\ref{algorithm:2}, we split the data into two halves and choose to estimate $\pi_T$ and $S$ on one half and estimate $\psi_0$ and $\beta_0$ on the other. This separates estimation errors of $\psi_0$ and $\beta_0$ from those of $\pi_T$ and $S$. 
The theoretical property of the Algorithm~\ref{algorithm:2} is investigated in Section~\ref{sec:theory}. As shown in Theorem~\ref{thm:consistency}, the convergence rate of $\widehat{\beta}$ is the same as that of $\widehat{\psi}$ which converges slower than $n^{-1/2}$. This is a consequence of using an mis-specified $\alpha_*( X)$ and shows that a subsequently efficiency augmentation procedure is necessary to improve the efficiency of the coefficient estimates.

\bigskip

\begin{minipage}{0.95\linewidth} 
\begin{algorithm}[H]\label{algorithm:2}
	\caption{Estimation of $\psi(  X^\top \beta)$ with {\em unknown} ${\pi}_T$ and ${S}$}
	\SetAlgoLined
	Randomly split data into two parts $\mathcal{I}^{(1)}$ and  $\mathcal{I}^{(2)}$\label{algo2:1} \;
	Estimate $\pi_T$ and $S$ on $\mathcal{I}^{(1)}$ and denote the estimators as $\hat{\pi}^{(1)}$ and $\hat{S}^{(1)}$\label{algo2:2} \;
    Obtain the estimates $\hat{  \xi}^{(2)}$ and $\hat{ \beta}^{(2)}$ on $\mathcal{I}^{(2)}$ by Algorithm~\ref{algorithm:1} with the estimated $\hat{\pi}^{(1)}$ and $\hat{S}^{(1)}$ from Step~\ref{algo2:2}. In particular,  $\hat{\psi}^{(2)}(\cdot)=\sum_{j=1}^d\hat{\xi}^{(2)}_jN_j(\cdot)$	\label{algo2:3}\;
	 Switch the order of $\mathcal{I}^{(1)}$ and $\mathcal{I}^{(2)}$, and repeat Steps~\ref{algo2:2} and \ref{algo2:3} to obtain $\hat{\psi}^{(1)}$ and $\hat{ \beta}^{(1)}$ \;
	 Aggregate to obtain	$
	\hat{ \beta}=\frac{1}{2}\sum_{s=1}^2 \hat{ \beta}^{(s)}$ and $\hat{\psi}=\frac{1}{2}\sum_{s=1}^2 \hat{\psi}^{(s)}. \label{algo2:4}$ 
\end{algorithm}
\end{minipage}

\subsection{The second step: one-step efficiency augmentation procedure using the efficient score}
\label{sec: inference}

In this section, we investigate an efficiency augmentation procedure for $\widehat{\beta}$ using the efficient score. In Section~\ref{sec: efficiency_theory}, we mentioned that the key element in using the efficient score requires evaluation of $\alpha_*(X)$, which depends on $V(X)$, $ {E}(V(X)S^2\mid X^\top\beta)$, and $ {E}(V(X)S^2 X\mid X^\top\beta)$. Suppose that we have an initial estimator $\widehat{\xi}$,  $\widehat{\beta}$, $\widehat{\pi}$, $\widehat{S}$, and estimators for the conditional variance terms $\sigma^2_T(X)$, we can obtain a plug-in estimator for $V(X)$. {We propose to estimate $\sigma^2_T(X)$ from outcome models for the $T=1$ and $T=-1$ groups respectively. For $T=1$, after obtaining the estimated outcome model $\widehat{\mu}_1(X)$, we can regress the squared-residual $(Y-\widehat{\mu}_1(X))^2$ on $X$ using kernel regression to estimate $\sigma_1^2(X)$. The $\sigma_{-1}^2(X)$ can be estimated similarly from the $T=-1$ group. }

Further, we can estimate $ {E}(V(X)S^2 X\mid X^\top\beta)$ and $ {E}(V(X)S^2\mid X^\top\beta)$ by fitting $\widehat{V}(X)\widehat{S}^2 X$ and $\widehat{V}(X)\widehat{S}$ on $ X^\top\widehat{\beta}$ using kernel regression.  An estimator for ${\alpha}_*(X)$ can be formed by plugging-in these components. As shown in the proof of Theorem~\ref{thm:asymptotic_normal}, the estimator $\widehat{\alpha}_*( X)$ converges to ${\alpha}_*( X)$ in certain rate.

Now write the efficient score in Theorem~\ref{thm:efficient_score} as
\begin{equation}\label{eq:score}
 {E}_n\left (\nabla \ell(  \xi,\beta;\pi_T,{S})\nabla\psi\left( X^\top\beta\right)\alpha_*( X)\right),
\end{equation} where 
$\ell( \xi,\beta;\pi_T, S)$ was defined in \eqref{loss_function2}. We propose a modified one-step estimator based on the first order expansion of \eqref{eq:score} as follows
\begin{eqnarray}\nonumber
&& {E}_n\left (\nabla \ell(\widehat{  \xi}, \widehat{\beta}; \widehat{\pi}_T, \widehat{S})\nabla\widehat{\psi}\left( X^\top\widehat{\beta}\right )\widehat{\alpha}_*( X)\right )+\\
&&\left\{ {E}_n\left (\nabla^2 \ell(\widehat{  \xi}, \widehat{\beta}; \widehat{\pi}_T, \widehat{S})\left\{\nabla\widehat{\psi}\left( X^\top\widehat{\beta}\right)\right\}^2 X\widehat{\alpha}_*( X)\right )+\lambda_n  I\right\}(\beta-\widehat{\beta})=0, \label{eq:one-step}
\end{eqnarray}
where $\lambda_n$ is a parameter and $  I$ is an identity matrix with dimension $p\times p$. Compared with the usual one-step estimator \citep{Vaart1999}, we have the extra term $\lambda_n  I$ in \eqref{eq:one-step} due to the constraint $\|\beta\|_2=1$. This constraint makes the matrix
\begin{equation*}
	 \Sigma_1 \equiv  {E}\left (\nabla^2 \ell(\psi_0( X^\top\beta_0) ;{\pi}_T, {S})\left\{\nabla \psi_0( X^\top\beta_0)\right\}^2 X\alpha_*^\top( X)\right )
	\end{equation*}  
not of full rank (the rank is $p-1$). As a result, its approximation 
\begin{equation*}
	\widehat{ \Sigma}_1 \equiv   {E}_n\left (\nabla^2 \ell(\widehat{  \xi}, \widehat{\beta}; \widehat{\pi}_T, \widehat{S})\left\{\nabla\widehat{\psi}\left( X^\top\widehat{\beta}\right)\right\}^2 X\widehat{\alpha}_*( X)\right )
\end{equation*}
may not be invertible numerically. As such, we add a small $\lambda_n$ to the diagonal. As we will show in Section~\ref{sec:theory}, with an appropriate choice of $\lambda_n$, the one-step efficiency augmentation estimator $\widetilde{\beta}$ defined as the solution of \eqref{eq:one-step} is asymptotically normal {and semiparametrically efficient},
\begin{eqnarray} 
\widetilde{\beta} = \widehat{\beta} + \left\{\widehat{ \Sigma}_1+\lambda_n  I\right\}^{-1} {E}_n\left (\nabla \ell(\widehat{  \xi}, \widehat{\beta}; \widehat{\pi}, \widehat{S})\nabla\widehat{\psi}\left( X^\top\widehat{\beta}\right )\widehat{\alpha}_*( X)\right ) \label{eq:solution to one-step}.
\end{eqnarray}
{The estimator of the asymptotic variance of $\widetilde{\beta}$ can be defined as 
\begin{equation}
\widetilde{ \Sigma}=\left\{\widehat{ \Sigma}_1+\lambda_n  I\right\}^{-1} {E}_n\left (\left\{\nabla \ell(\widehat{  \xi}, \widehat{\beta}; \widehat{\pi}, \widehat{S})\nabla\widehat{\psi}\left( X^\top\widehat{\beta}\right )\right\}^2\widehat{\alpha}_*( X)\widehat{\alpha}_*^\top( X)\right )\left\{\widehat{ \Sigma}_1+\lambda_n  I\right\}^{-1}\label{eq:variance of one-step}.
\end{equation}}

{To get initial estimators $\widehat{ \xi}$,  $\widehat{\beta}$, $\widehat{\pi}$, and $\widehat{S}$, we can split the dataset into three parts. On the first part, we fit $\pi$, $\mu_1$, and $\mu_{-1}$; on the second part, we utilize the proposed doubly robust loss with estimated $\pi$ and $S=\mu_1+\mu_{-1}$ to get initial estimators $\widehat{ \xi}$ and $\widehat{\beta}$; finally, we improve the estimator using the efficiency augmentation procedure on the third part. To make up efficiency loss of sample-splittings, we rotate the role of each part and extend the cross-fitting algorithm. Details of this algorithm is listed in Algorithm \ref {algorithm:3}.}

\bigskip

\begin{minipage}{0.95\linewidth} 
\begin{algorithm}[H]\label{algorithm:3}
	\caption{Semiparametrically efficient estimation  of $ \beta$ based on sample-splitting}
	\SetAlgoLined
Randomly split data into three parts $\mathcal{I}^{(1)}$,  $\mathcal{I}^{(2)}$, and $\mathcal{I}^{(3)}$ \label{algo3:1} \;	

Use $\mathcal{I}^{(1)}$ to obtain estimators $\hat{\pi}^{(1)}$ and $\hat{S}^{(1)}$; use $\mathcal{I}^{(2)}$ to obtain estimators $\hat{\xi}^{(2)}$ and $\hat{ \beta}^{(2)}$\label{algo3:2} \;

Obtain $\hat{\psi}^{(2)}$, $\hat{V}^{(2)}(  X),$ and $\hat{\alpha}_*^{(1,2)}(  X)$ using the estimators in Step \ref{algo3:2}. In obtaining $\hat{\alpha}_*^{(1,2)}(  X)$, we estimate $\mathbb{E}[V(  X)S^2  X|  X^\top \beta]$, $\mathbb{E}[V(  X)S^2|  X^\top \beta]$ using $\mathcal{I}^{(2)}$ based on Gaussian kernel regressions of bandwith $n^{-1/5}$. Likewise, we estimate $\mathbb{E}[\epsilon^2\,|\,T=1,  X]$ and $\mathbb{E}[\epsilon^2\,|\,T=-1,  X]$ using $\mathcal{I}^{(2)}$ based on Gaussian kernel regressions of bandwith $n^{-1/(2p+1)}$\label{algo3:3}\;
	 
Calculating $\tilde{ \beta}^{(3)}$ and $\tilde{\Sigma}^{(3)}$ using \eqref{eq:solution to one-step} and \eqref{eq:variance of one-step} where the empirical expectation is defined by data in $\mathcal{I}^{(3)}$ and the unknown parts are based on $\hat{  \xi}^{(2)}, \hat{ \beta}^{(2)}; \hat{\pi}^{(1)}, \hat{S}^{(1)}, \hat{\psi}^{(2)}$ and $\hat{\alpha}_*^{(1,2)}(  X)$\label{algo3:4}\;	 

Switch the order of $\mathcal{I}^{(1)}$,  $\mathcal{I}^{(2)}$, and $\mathcal{I}^{(3)}$ from Step~\ref{algo3:1}, and repeat Steps~\ref{algo3:2} and~\ref{algo3:4} to obtain $\tilde{ \beta}^{(s)}$, $\hat{\psi}^{(s)}$, and $\tilde{\Sigma}^{(s)}$ for $s=1, 2, 3$ \;
 
  Aggregate to obtain $\tilde{ \beta}=\frac{1}{3}\sum_{s=1}^3\tilde{ \beta}^{(s)}$, $\hat{\psi}=\frac{1}{3}\sum_{s=1}^3 \hat{\psi}^{(s)}$,  $\hat{\theta}(  X)=\hat{\psi}(  X^\top\tilde{ \beta})$, and   $\tilde{ \Sigma}=\frac{1}{3}\sum_{s=1}^3\tilde{ \Sigma}^{(s)}$. 	\label{algo3:5}
\end{algorithm}
\end{minipage}


\section{Theoretical properties}
\label{sec:theory}

In this section, we present the theoretical properties of our proposed estimation procedures in Algorithms \ref{algorithm:1} -- \ref{algorithm:3}. We will present results for Algorithms~\ref{algorithm:1} and~\ref{algorithm:2} in Theorem  \ref{thm:consistency} and  
results for Algorithm~\ref{algorithm:3} in Theorem \ref{thm:asymptotic_normal}.    
 {Let $k$ be the smoothness parameter of the wrapper function $\psi_0$, which indicates that $\psi_0$ has at least $k$th bounded derivative over the support of $ X^\top\beta_0$.} The conditions that we need for our theoretical results are as follows.

\begin{enumerate}[label=(C.\arabic*)]
\item \label{Con:1} $\psi_0$ is non-decreasing with the smoothness parameter $k>3$, and  $\psi_0^{'}( X^\top\beta_0)> 0$ for some $X$.
	\item \label{Con:2} $\beta_0$ lies on the unit ball in $ {R}^p$ and $ X$ has a compact support. $ {E}\left ( X X^\top\mid X^\top\beta_0\right)$ is positive definite and $ {E}\left( X\mid X^\top\beta\right )$ has at least $k$th bounded and continuous derivative at $\beta_0$.	
	\item \label{Con:3}${E}\left(\psi_0\left( X^\top\beta_0\right)\mid X^\top\beta\right)$ is continuously differentiable in $\beta$ and 
	\begin{equation*}
	 {E}\left( \bigg\{\nabla {E}\left(\psi_0\left( X^\top\beta_0\right)\mid X^\top\beta\right) \mid {}_{\beta=\beta_0} \bigg\}^{\otimes 2} \right)>0,
	\end{equation*}  
	where for any vector $a$, ${a}^{\otimes 2}=aa^\top$.
	\end{enumerate}

For the estimated propensity and outcome models, we also require the following conditions on the convergence rate of these estimated nuisance parameters. Note that these conditions  are rather mild and many machine learning methods satisfy these conditions. Due to the (extended) cross-splitting procedures in  Algorithms~\ref{algorithm:2} and~\ref{algorithm:3}, these conditions and corresponding conclusions are expected to hold for split datasets as their sample sizes are in the same order of the whole data set. 

\begin{enumerate}[label=(C.\arabic*)]
	\setcounter{enumi}{3}
	\item \label{Con:4} Suppose that for some $\alpha,\beta>0$, $\|\widehat{\pi}-\pi\|_{\infty}=O_p(n^{-\alpha})$ and $\|\widehat{\mu}_1-\mu_1\|_{\infty}+\|\widehat{\mu}_{-1}-\mu_{-1}\|_{\infty}=O_p(n^{-\beta})$. Then we require that ${\alpha+\beta} > {1/2}$.
	\item \label{Con:5} There  exists two constants $\pi_{\min}$ and $\pi_{\max}$ satisfying $0<  \pi_{\min}\leq \pi ( X), \widehat{\pi}( X) \leq \pi_{\max} <1$ for all $ X$.
\end{enumerate}
Condition~\ref{Con:4} requires consistent estimators for $\pi$ and outcome models ( or $S=\mu_1+\mu_{-1}$), but the convergence rates can be nearly as slow as $n^{-1/4}$. In addition, the condition mitigates the requirement of the convergence rate on $S$ when we have an estimator of $\pi$ with relatively fast convergence rate and vice versa. For example, if $\pi$ can be estimated by a logistic regression and the convergence rate of $\widehat{\pi}$ is $O_p(n^{-1/2})$, then, we only require the convergence rate on $\widehat{S}$ to be $o_p(1)$. This indicates that $S$ can be obtained by many nonparametric methods. Further, if we use the kernel regression with the optimal bandwidth, the convergence rate of $\widehat{S}$ is as fast as $\left(n^{-1}\log n\right)^{l/(2l+p)}$ for $l$th order Lipschitz continuous $ {E}(Y_a\mid X)$ \citep{stone1982}. If both nuisance parts are estimated by the kernel regression and have the same order of smoothness, we have $n^{-\alpha-\beta}\simeq \left(n^{-1}\log n\right)^{2l/(2l+p)}$ which is much faster than $n^{-1/2}$ when $p<2l$. In addition to Condition~\ref{Con:4}, we also require another condition on $\max\left\{n^{-\alpha}, n^{-\beta}\right\}$. This condition depends on the {smoothness} of $\psi_0$. Therefore, it is introduced with the Theorems. Condition ~\ref{Con:5} is commonly known as the positivity assumption for the propensity score \citep{Rosenbaum1983}.

Because Algorithm~\ref{algorithm:2} utilizes Algorithm~\ref{algorithm:1} on split datasets, the convergence rates of the estimators obtained in Algorithm~\ref{algorithm:1} also apply to  Algorithm~\ref{algorithm:2} under the same conditions. Denote the final estimators of Algorithm~\ref{algorithm:1} as $\widehat{\beta}_n$ and $\widehat{\psi}_n$. Under Conditions~\ref{Con:1} -~\ref{Con:5}, the following theorem establishes the consistency and convergence rates of  $(\widehat{\beta}_n, \widehat{\psi}_n)$. 

\begin{theorem}\label{thm:consistency}
	Assume $K_n=C_1n^{\gamma}$ for some positive constants $C_1$ with with $\gamma < 1/10$. Then under Conditions~\ref{Con:1} -~\ref{Con:5}, for any pre-specified $\nu$ with $0<\nu\leq 1/2$, we have
	\begin{equation*}
	\|\widehat{\beta}_n-\beta_0\|^2_2+\|\widehat{\psi}_n-\psi_0\|_{L_2}^2=O_p\left(n^{-1+\nu+\gamma}\right)+O(n^{-2k\gamma}).
	\end{equation*}
\end{theorem}

Theorem~\ref{thm:consistency} justifies the consistency of Algorithms~\ref{algorithm:1} and~\ref{algorithm:2}. The resulting convergence rate has two terms. The first term $O_p\left(n^{-1+\nu+\gamma}\right)$ represents the variance of the estimation. When $\gamma$ is large, $-1+\nu+\gamma$ becomes larger. The second term measures the bias of the estimation, which decreases with $\gamma$ increases. It is possible to choose an optimal $\gamma$ based on  convergence rates of the nuisance parameters and the smoothness parameter $k$. When ${(1-\nu)}/{(2k+1)}\leq 1/10$, we can set $\gamma={(1-\nu)}/{(2k+1)}$, and it leads to a convergence rate of $O_p\left(n^{-{2k(1-\nu)}/{(2k+1)}}\right)$. When ${1}/{(2k+1)}\geq 1/10$, we can only set $\gamma=1/10$. 

The following Theorem provides the asymptotic normality of the one-step efficient estimator $\widetilde{\beta}$ resulted from Algorithm~\ref{algorithm:3}. Denote $\mu_{\beta}(t)= {E}(V(X)S^2 X\mid X^\top\beta=t)$ and $\xi_{\beta}(t)= {E}(V(X)S^2\mid X^\top\beta=t)$. Let $\hbar\equiv n^{-1/2\min\left\{\alpha,\beta\right\}}+n^{-1/4+\nu/4+\gamma}+n^{-(k/2-3/4)\gamma}+\left(n/\log n\right)^{1/(p+4)}$.

\begin{theorem}\label{thm:asymptotic_normal}
	Under Conditions~\ref{Con:1} -~\ref{Con:5}, we further assume 1) $\mu_{\beta}(t)$ and $\xi_{\beta}(t)$ have bounded derivatives w.r.t $\beta$ and $t$; 2) $K_n=C_1n^{\gamma}$ for some positive constants $C_1$ with $\gamma<1/10$; and 3) $n^{1/2}\hbar(n^{-1/2+\nu/2+2\gamma}+n^{-(k-3/2)\gamma})\to 0$. If we set $\lambda_n=\hbar$, then $n^{1/2}\left(\widetilde{\beta}-\beta_0\right)$ converges in distribution to a mean-zero normal distribution with covariance matrix $ \Sigma=\Sigma_1^{+} \Sigma_2 \Sigma_1^{+}$, where $ \Sigma_1^{+}$ is the pseudo inverse of $\Sigma_1$, and the formula of $\Sigma_1$ and $\Sigma_2$ can be found in the supplementary material.
\end{theorem}

In Theorem~\ref{thm:asymptotic_normal}, the asymptotic variance comes from a degenerated Gaussian distribution because $\beta_0^\top  \Sigma_1\beta_0=0$ and $\beta_0^\top  \Sigma_2\beta_0=0$. The result in Theorem~\ref{thm:asymptotic_normal} shows that the one-step estimator is asymptotically normal if $\lambda_n$ is set as $\hbar$ and satisfies that $n^{1/2}\hbar(n^{-1/2+\nu/2+2\gamma}+n^{-(k-3/2)\gamma})\to 0$. The convergence rate of $\hbar$ cannot be faster than $n^{-1/2\min\left\{\alpha,\beta\right\}}$. When $\alpha$ and $\beta$ are small, the smoothness parameter $k$ of $\psi_0$ needs to be large with a fixed $\gamma$ such that $n^{1/2}\hbar(n^{-1/2+\nu/2+2\gamma}+n^{-(k-3/2)\gamma})\to 0$. The proof of Theorem~\ref{thm:asymptotic_normal} also implies that the variance estimator in Algorithm~\ref{algorithm:3} is consistent.

\section{Simulation}
\label{sec:sim}

In this section, we provide simulation results for our proposed method based on Algorithms~\ref{algorithm:2} and~\ref{algorithm:3}. {In these simulations, we compare with the Q-learning method \citep{qian2011}, the single index model on the absolute difference function $\Delta( X)$ ($\Delta$-learning) proposed in \citep{song2017} and \citep{Liang2020}, and the EARL method \citep{zhao2019}.} Q-learning is a regression-based method, which posits parametric assumptions on the outcome models. $\Delta$-learning is extended in \citep{Liang2020} to incorporate a doubly robust step for improving efficiency and safeguard model-misspecification. The EARL method  is an efficiency augmented approach that extends the O-learning method of \cite{Zhao2012}.

In these comparisons, we focus on two metrics evaluated on a testing data set. The first metric is rank correlation between the estimate $\widehat \theta$ and the true $\theta$. The second metric is the value function defined as
${V}(d( X))= {E}(Y_{d( X)})$ where  $d(X)=\mathrm{sign}\{\widehat{\theta}\}$.  Given a testing dataset, we can empirically estimate this value function by
$\widehat{{V}}(d( X))= {E}_n(Y_{d(X)}).$
 In many practical settings, the rank correlation is of more interests than the value function because the former facilitates prioritization of patient recommendation. In addition to these comparisons, we also provide coverages of the interval estimation from Algorithm~\ref{algorithm:3}.

For our proposed approach, we obtained $\widehat{\theta}$ using the output of Algorithm~\ref{algorithm:3}. Kernel regressions were used to estimate the propensity and outcome models. For the Q-learning approach, we estimated ${\mu}_1( X)$ and  ${\mu}_{-1}( X)$ first from  linear models with all the covariates and the interaction terms between the covariates and treatment. Then we constructed a plug-in estimator for $\theta$. {For the $\Delta$-learning approach, we posit a single index model on the absolute difference $\Delta( X)={\mu}_1( X)-{\mu}_{-1}( X)$ following \citet{song2017}. To obtain the relative contrast $\theta(X)$, we estimate $S(X)$ by using a single index model on $Y$ because $E(Y/\pi_T( X)\mid X) = S(X)$.  For the EARL method which focuses on classification, the optimal individualized treatment rule which minimizes the value function ${V}(d(X))$ can be simplified to a linear decision rule when $\psi_0$ is increasing. As such, for EARL, we consider  only the class of linear decision rules and use its resulting estimate $ X^\top\widehat{\beta}_{\mathrm{EARL}}$ as an estimator for $\theta$.} Kernel regressions were used to estimate the propensity scores in all methods except in Q-learning when they are not needed. Kernel regressions were also used for outcome models as required by the Q-learning  and by the augmentation part of the EARL method.

To generate outcomes, we consider the following two settings: O1) All covariates $ X$ are generated from independent  uniform distribution on $[-0.5,0.5]$. We set
$$\theta=\log \left(\{1.5+\Phi( X^\top\beta_0)\}/\{2.5-\Phi( X^\top\beta_0)\}\right),$$ where $\beta_0=(1,-1,1,-1)^\top$ and $\Phi(\cdot)$ is the cumulative distribution function of standard normal $N(0,1)$. We further set $S=\| X\|_2$. Then from the relationship $\mu_{1}( X) =e^\theta S/(e^\theta + 1)$ and $\mu_{-1}( X) = \mu_{1}( X)/e^{\theta}$, we generate $Y_1=\mu_{1}( X)+ N(0,0.01)$ and $Y_{-1}=\mu_{-1}( X)+ N(0,0.01)$; O2)
	All covariates $ X$ are generated from independent  uniform distribution on $[-1,1]$. Here $$\theta=\log \left(\{0.5+\Phi(0.8 X^\top\beta_0)\}/\{1.5-\Phi(0.8 X^\top\beta_0)\}\right),$$ with $\beta_0=(-1,-1,1,-1)^\top$. We set $S=\exp\{0.8 X^\top\beta_0\}+1$. We generate $Y_1$ and $Y_{-1}$ according to Poisson distributions with mean $\mu_{1}( X)$ and $\mu_{-1}( X)$. More simulation settings with correlated variables are available in the supplementary materials.

For the propensity model, we consider the following two settings:
PS~1) $ {P}(T=1| X)=e^{ X^\top\beta_{\pi}}/(1+e^{ X^\top\beta_{\pi}})$, {where $\beta_{\pi}=(0.2,-0.2,-0.2,0.2)^\top$}; PS~2) $${P}(T=1| X)=\exp\{0.2(X_1^2+X_2^2+X_1X_2)\}/\left\{1+\exp\{0.2(X_1^2+X_2^2+X_1X_2)\}\right\}.$$
Therefore we consider four simulation scenarios O1)+PS1), O2)+PS1), O1)+PS2), and O2)+PS2). In each simulation, the sample size varied from $800$, $1000$, to $1500$ for training samples. To evaluate different methods, we generate a testing dataset with sample size of $10^5$. Each simulation is repeated for $500$ times. 

\begin{figure}
	\centering
	\includegraphics[width=0.95\linewidth]{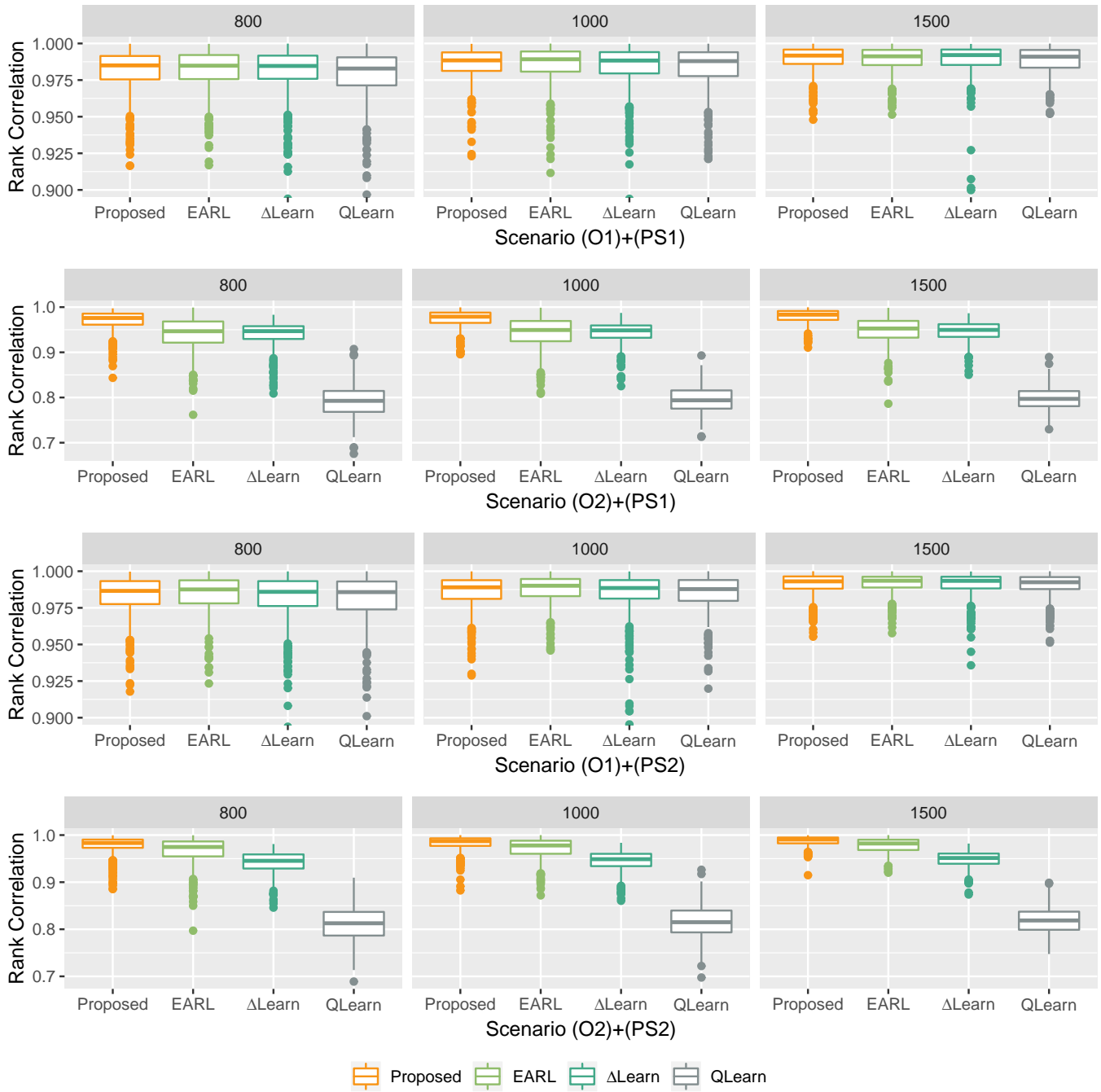}
	\caption{Rank correlation comparison in simulated data.}
	\label{fig:fig1}
\end{figure}

\begin{figure}
	\centering
	\includegraphics[width=0.95\linewidth]{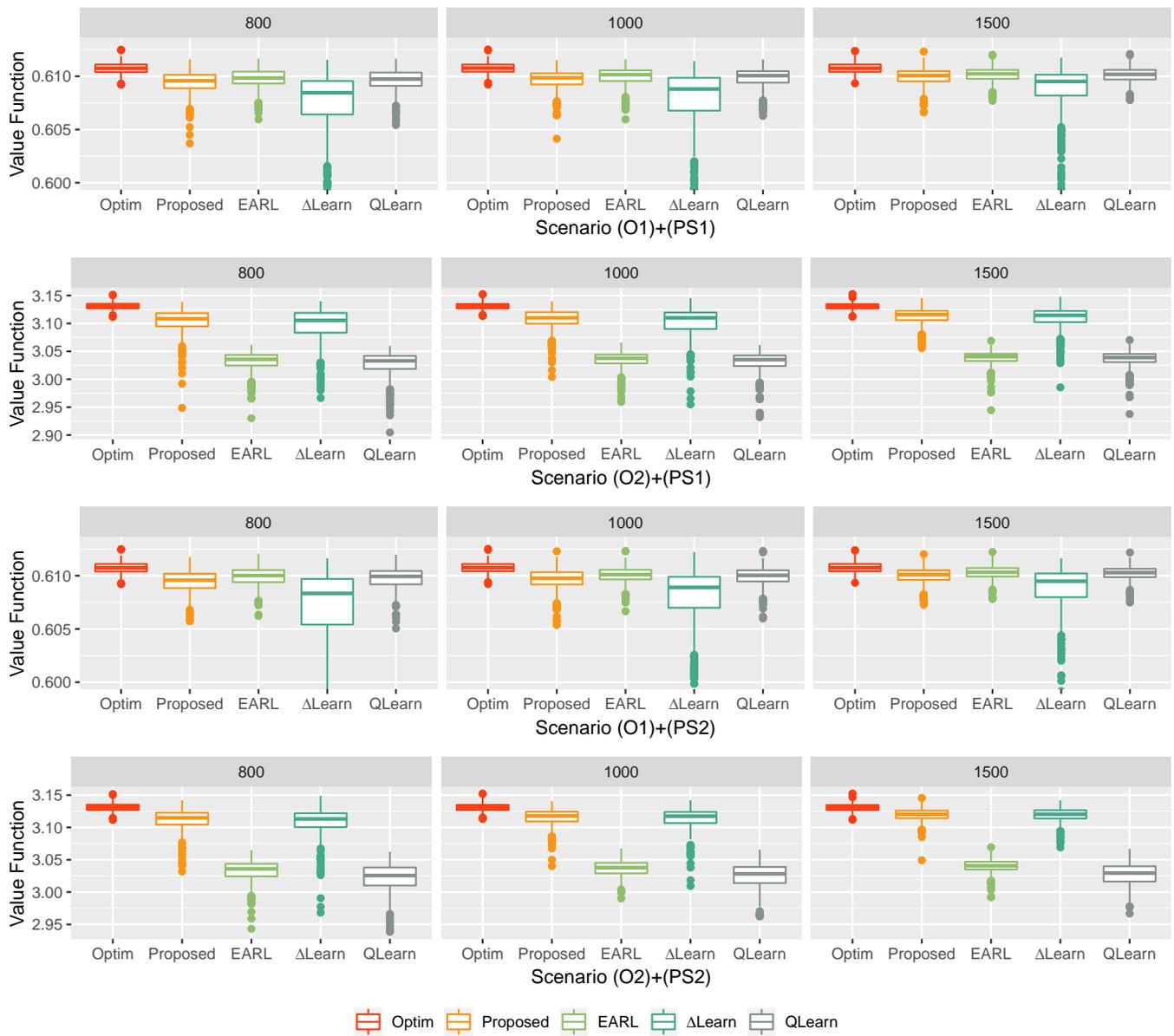}
	\caption{Value function comparison in simulated data.}
	\label{fig:fig2}
\end{figure}

Figures~\ref{fig:fig1} and~\ref{fig:fig2} show the results for the rank correlation and the value function, respectively. For comparison, we also plotted the oracle value of the value function (labeled as `Optim') in Figure~\ref{fig:fig2}. We see that our method outperforms Q-learning in all scenarios in term of the rank correlation. EARL also achieves higher rank correlation and value function compared with the Q-learning. {$\Delta$-learning achieves higher rank correlation but lower value compared with the Q-learning.} Comparing with EARL and C-learning, we find that our proposed method preforms comparably in Setting~O1), but much better in Setting~O2). These results show that our proposed method preform consistently well under our simulation scenarios. In terms of the inference results, Table~\ref{tab:coef} shows the coverages of the interval estimation from Algorithm~\ref{algorithm:3}. They appear to work very well.

\begin{table}
	\caption{Coverages of the $95\%$ confidence intervals from our proposed approach.}
	\label{tab:coef}
	\begin{center}
		\begin{tabular}{cccccc}
			\toprule
			& & \multicolumn{4}{c}{Scenarios} \\
			\cline{3-6}
			Sample Size& & O1+PS1 &  O2+PS1 & O1+PS2 & O2+PS2   \\
			\hline\\[-2.5ex]
			\multirow{4}{*}{800}&$\hat{\beta}_1$ &0.974 & 0.930& 0.956 & 0.952  \\
			& $\hat{\beta}_2$ & 0.962& 0.942& 0.964&0.942 \\
			& $\hat{\beta}_3$ & 0.948 & 0.930 & 0.950&0.946\\
			& $\hat{\beta}_4$& 0.952&0.942 &0.962 & 0.944 \\
			\hline\\[-2ex]
			\multirow{4}{*}{1000}&$\hat{\beta}_1$ & 0.964 & 0.914 & 0.976 & 0.960 \\
			& $\hat{\beta}_2$ & 0.972 & 0.948 & 0.974&0.958 \\
			& $\hat{\beta}_3$ & 0.946 & 0.936 & 0.950&0.962\\
			& $\hat{\beta}_4$& 0.956 & 0.930 &  0.964&0.954 \\
			\hline\\[-2ex]
			\multirow{4}{*}{1500}&$\hat{\beta}_1$ & 0.964 & 0.930 & 0.988 & 0.954  \\
			& $\hat{\beta}_2$ & 0.964 & 0.930 & 0.968&0.965 \\
			& $\hat{\beta}_3$ & 0.962 & 0.926 & 0.980&0.960\\
			& $\hat{\beta}_4$& 0.954 & 0.934 & 0.974 &0.960 \\
			\bottomrule
		\end{tabular}
	\end{center}
\end{table}

\section{Real data example}
\label{sec:real data}

In this section, we compare different approaches in analyzing data   from a study of the intervention effects of phone counseling versus usual care on the adherence to mammography screening guidelines \citep{champion2007effect}. Participants included female subjects who did not follow the  mammogram guidelines in the year before entering the study. One primary outcome of the study was whether the participant took mammography screening during a 21-month follow-up post-baseline period. Among 530 subjects, 259 participants received the phone intervention and 271 the usual care. {Each participant has 18 covarites available including demographics, number of years had a mammogram, recommendation by a doctor or nurse, and mammography screening and breast cancer belief and knowledge scales.}

We randomly split the whole dataset into training and testing datasets with a sample size ratio of 4:1. On the training dataset, our proposed method, Q-learning, $\Delta$-learning, and EARL were implemented, similar to our simulation study.  On the testing datasets, we calculate two following metrics to compare different methods. The whole procedure is repeated $1000$ times. 

\begin{figure}
	\centering
	\includegraphics[width=0.8\linewidth]{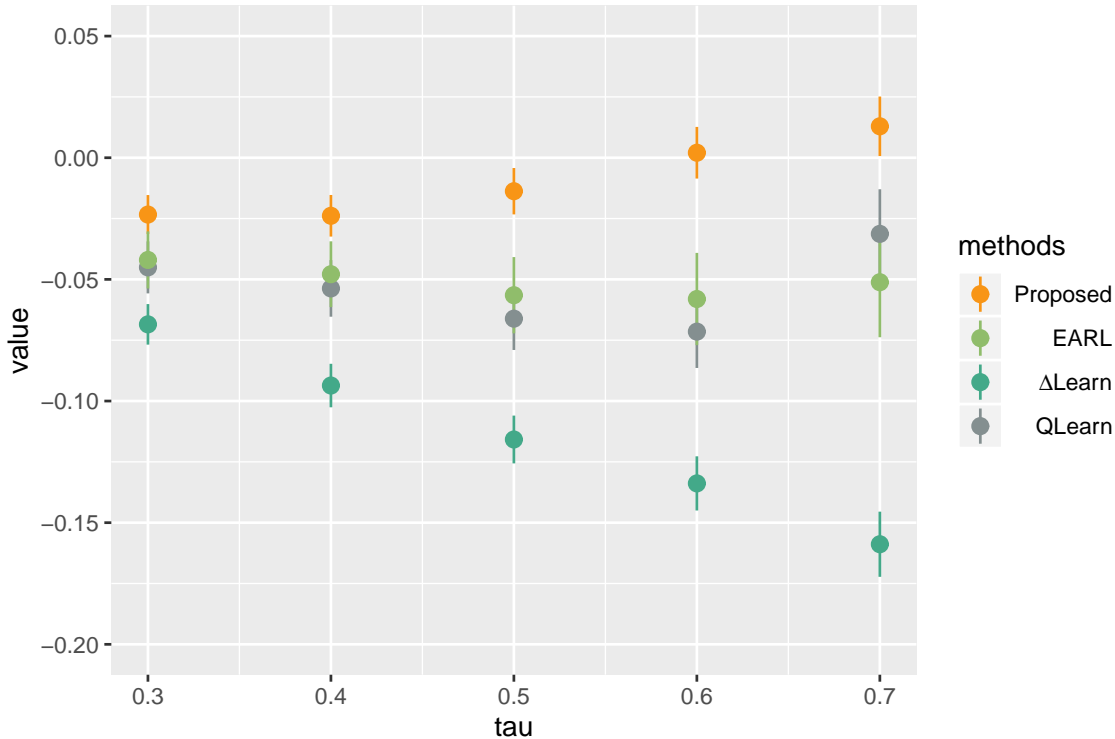}
	\caption{Comparison of the ARC at different levels of $\tau$.}
	\label{fig:fig3}
\end{figure}

The first metric is the value function for a given ITR $d(  X)$ as in the simulation study. The second metric is related to the ranking of $\hat{\theta}$. However, because the true $\theta$ is unknown, we propose the following averaged relative change (ARC) as our second metric for comparison. Specifically, for any $\tau\in(0,1)$, we define
\begin{equation}
ARC(\hat{\theta};\tau)=\dfrac{\mathbb{E}_n\left[\dfrac{TY}{\pi_T}I\big\{  X: \hat{\theta}(  X)\geq \hat{\theta}_\tau\big\}\right] }{\mathbb{E}_n\left[\dfrac{Y}{\pi_T}I\big\{  X: \hat{\theta}(  X)\geq \hat{\theta}_\tau\big\}\right]},
\end{equation}
where $\hat{\theta}_\tau$ is the $\tau$th-quantile of $\hat{\theta}$ evaluated on a testing data set. It is easy to see that the ARC can be written as the ratio of the average treatment effect, $Y_1 - Y_{-1}$, on $\{  X: \hat{\theta}(  X)\geq \hat{\theta}_\tau\}$ and the average main effect, $Y_1 + Y_{-1}$, on $\{  X: \hat{\theta}(  X)\geq \hat{\theta}_\tau\}$. Therefore if $\hat{\theta}$ possesses good ranking ability, we would expect $ARC(\hat{\theta};\tau)$ to be non-decreasing as a function of $\tau$. 
 Table~\ref{tab:value_data} and Figure~\ref{fig:fig3} show that our proposed method performs the best based on the above two metrics. In Figure, we only plotted $ARC(\hat{\theta};\tau)$ for $\tau=0.3, 0.4, \cdots, 0.7$. We also implemented our proposed method on the entire dataset to identify the important covariates shown in Table~\ref{tab:coef_data}. 


\renewcommand{\arraystretch}{1}
\begin{table}
	\caption{Comparison of the value function results in the real data example summarized over $1000$ cross-validated repeats.}
	\label{tab:value_data}
	\begin{center}
		\begin{tabular}{ccccc}
			\toprule\\[-3ex]
		 & \multicolumn{4}{c}{\bf Method}    \\
		 \cline{2-5}
		 &	Proposed& EARL  & $\Delta$Learn &  QLearn\\
			\hline\\[-2.5ex]
Value &  0.4906 & 0.4845& 0.4526 & 0.4805 \\
 SE     & 0.0022 &  0.0012& 0.0017 & 0.0021 \\
			\bottomrule
		\end{tabular}
	\end{center}
\end{table}



\begin{table}[h]
	\caption{Estimated coefficients with P-value $< 0.05$ in the real data example}
	\label{tab:coef_data}
	\begin{center}
		\begin{tabular}{ccc}
			\toprule
			{\bf Variable} & Coef & P-value   \\
			\hline\\[-2.5ex]
			Currently working for pay (Yes) & 0.496 & $2.72\times 10^{-8}$\\
			Benefits scale score & 0.248 & $8.80\times 10^{-7}$\\
			Number of years had a mammogram & 0.172 & $1.06\times 10^{-6}$\\
			Low household income (Yes) & 0.365 & $1.71\times 10^{-5}$\\
			Caucasian (Yes) & -0.265 & $1.03\times 10^{-3}$\\
			More than a High School degree (Yes) & 0.300 & $2.05\times 10^{-3}$\\
			Baseline stage of mammography screening behavior (contemplation) & 0.241 & $2.31\times 10^{-3}$\\
			Fear scale score & -0.153 & $4.76\times 10^{-3}$\\
			Family history of breast cancer (Yes) & 0.403 & $6.21\times 10^{-3}$\\
			Fatalism scale score & 0.085 & $4.12\times 10^{-2}$\\
			Susceptibility scale score & 0.126 & $4.54\times 10^{-2}$\\
			\bottomrule
		\end{tabular}
	\end{center}
\end{table}

\section{Conclusion}
\label{sec:diss}

Most existing literature on individualized treatment rule targets a contrast based on treatment effect difference as benefit. However, there can be settings where a relative difference may better represent such benefit. In this work, we proposed to estimate treatment benefits using relative contrast functions. We showed that relative contrasts can be used for treatment recommendation as well as priority scores for  receiving different treatments. 

Because all relative contrasts are monotonic transformations of each other, we posited a semiparametric single index model for a particular relative contrast. Along with semiparametrically efficient estimation of the index parameters, we also provided asymptotic distributional results that can be used for inference even when nuisance parameters are estimated non-parametrically.  

{Another related theoretical issue is inference for the optimal value  \citep{chakraborty2010inference, chakraborty2014inference, luedtke2016statistical, zhang2012robust, goldberg2014comment, laber2014rejoinder}.  As discussed in our introduction, our proposed estimators can be used for treatment recommendation. Given the estimated $\widehat{\psi}( X^\top\widetilde{\beta})$, the estimated decision rule can be defined as $\mathrm{sign}\{\widehat{\psi}( X^\top\widetilde{\beta})\}$. To infer the value under such estimated decision rule, existing methods can be applied due to the relationship between $\theta( X)$ and $\Delta( X)$. In particular, the value function inference procedure proposed by \cite{shi2020breaking} should be applicable even their results are based on estimated $\Delta( X)$. The conditions in \cite{shi2020breaking} are very general and we can directly replace the estimator of $\Delta( X)$ with $\widehat{\psi}( X^\top\widetilde{\beta})$ to yield a valid inference procedure similar to  \cite{shi2020breaking}.}

One important future work is to extend our approach to accommodate high dimensional covariates, for example by adopting the de-correlated score proposed in \citet{Ning2017}. To circumvent the difficulty of enforcing the $L_2$-norm constraint, one can replace it by assuming that a model like $\psi_0(1+ X^\top\beta_0)$, where $\psi_0$ is a monotone function and $\psi_0(0)=0$. 

Another extension of our approach is to allow a non-monotone $\psi_0$. This extension is computationally straightforward by simply removal of the constraint on $ \xi$. However, it is possible that we will require the nuisance parameter convergence rates to be faster than what is required in our work. It may also be hard to interpret $\beta_0$ when $\psi_0$ is non-monotonic.

\eject 

\newpage 
\begin{center}
	{\large\bf Supplemental Materials}
\end{center}

\begin{description}
	
	\item[] \hspace{.65cm} Proofs of all theorems and additional simulation results are contained in the online supplemental materials. 
	
\end{description}

\singlespacing
\bibliographystyle{apalike}
\bibliography{ref.bib}

\begin{thebibliography}{}

\bibitem[Carroll et~al., 1997]{Carroll1997backfitting}
Carroll, R.~J., Fan, J., Gijbels, I., and Wand, M.~P. (1997).
\newblock Generalized partially linear single-index models.
\newblock {\em J. Am. Stat. Assoc.}, 92(438):477--489.

\bibitem[Chakraborty et~al., 2014]{chakraborty2014inference}
Chakraborty, B., Laber, E.~B., and Zhao, Y.-Q. (2014).
\newblock Inference about the expected performance of a data-driven dynamic
  treatment regime.
\newblock {\em Clin. Trials}, 11(4):408--417.

\bibitem[Chakraborty et~al., 2010]{chakraborty2010inference}
Chakraborty, B., Murphy, S., and Strecher, V. (2010).
\newblock Inference for non-regular parameters in optimal dynamic treatment
  regimes.
\newblock {\em Stat. Methods Med. Res.}, 19(3):317--343.

\bibitem[Champion et~al., 2007]{champion2007effect}
Champion, V., Skinner, C.~S., Hui, S., Monahan, P., Juliar, B., Daggy, J., and
  Menon, U. (2007).
\newblock The effect of telephone versus print tailoring for mammography
  adherence.
\newblock {\em Patient Educ. Couns.}, 65(3):416--423.

\bibitem[Chen et~al., 2017]{Chen2017}
Chen, S., Tian, L., Cai, T., and Yu, M. (2017).
\newblock A general statistical framework for subgroup identification and
  comparative treatment scoring.
\newblock {\em Biometrics}, 73(4):1199--1209.

\bibitem[Chernozhukov et~al., 2018]{victor2018}
Chernozhukov, V., Chetverikov, D., Demirer, M., Duflo, E., Hansen, C., Newey,
  W., and Robins, J. (2018).
\newblock Double/debiased machine learning for treatment and structural
  parameters.
\newblock {\em Econom. J.}, 21(1):C1--C68.

\bibitem[Dayan and Watkins, 1992]{Watkins1992}
Dayan, P. and Watkins, C. (1992).
\newblock Q-learning.
\newblock {\em Mach. Learn.}, 8(3):279--292.

\bibitem[Goldberg et~al., 2014]{goldberg2014comment}
Goldberg, Y., Song, R., Zeng, D., and Kosorok, M.~R. (2014).
\newblock Comment on “dynamic treatment regimes: Technical challenges and
  applications”.
\newblock {\em Electron. J. Statist.}, 8(1):1290.

\bibitem[{Guo} et~al., 2018]{wenchuan2018}
{Guo}, W., {Zhou}, X.-h., and {Ma}, S. (2018).
\newblock {Optimal treatment selection using the covariate-specific treatment
  effect curve with high-dimensional covariates}.
\newblock {\em arXiv e-prints}, page arXiv:1812.10018.

\bibitem[Imbens and Rubin, 2015]{Imbens2015}
Imbens, G.~W. and Rubin, D.~B. (2015).
\newblock {\em Causal Inference: For Statistics, Social, and Biomedical
  Sciences an Introduction}.
\newblock Cambridge Press.

\bibitem[Laber et~al., 2014a]{Laber2014}
Laber, E.~B., Linn, K.~A., and Stefanski, L.~A. (2014a).
\newblock {Interactive model building for Q-learning}.
\newblock {\em Biometrika}, 101(4):831--847.

\bibitem[Laber et~al., 2014b]{laber2014rejoinder}
Laber, E.~B., Lizotte, D.~J., Qian, M., Pelham, W.~E., and Murphy, S.~A.
  (2014b).
\newblock Dynamic treatment regimes: technical challenges and applications.
\newblock {\em Electron. J. Statist.}, 8(1):1225.

\bibitem[Leitenstorfer and Tutz, 2006]{leitenstorfer2006}
Leitenstorfer, F. and Tutz, G. (2006).
\newblock {Generalized monotonic regression based on B-splines with an
  application to air pollution data}.
\newblock {\em Biostatistics}, 8(3):654--673.

\bibitem[Liang et~al., 2010]{Liang2010profile}
Liang, H., Liu, X., Li, R., and Tsai, C.-L. (2010).
\newblock Estimation and testing for partially linear single-index models.
\newblock {\em Ann. Statist.}, 38(6):3811.

\bibitem[Liang and Yu, 2020]{Liang2020}
Liang, M. and Yu, M. (2020).
\newblock A semiparametric approach to model effect modification.
\newblock {\em J. Am. Stat. Assoc.}, 0(0):1--13.

\bibitem[Lu et~al., 2013]{lu2013}
Lu, W., Zhang, H.~H., and Zeng, D. (2013).
\newblock Variable selection for optimal treatment decision.
\newblock {\em Stat. Methods Med. Res.}, 22(5):493--504.

\bibitem[Luedtke and Van Der~Laan, 2016]{luedtke2016statistical}
Luedtke, A.~R. and Van Der~Laan, M.~J. (2016).
\newblock Statistical inference for the mean outcome under a possibly
  non-unique optimal treatment strategy.
\newblock {\em Ann. Statist.}, 44(2):713.

\bibitem[Ma and He, 2016]{Ma2016}
Ma, S. and He, X. (2016).
\newblock Inference for single-index quantile regression models with profile
  optimization.
\newblock {\em Ann. Statist.}, 44:1234--68.

\bibitem[Murphy, 2003]{murphy2003}
Murphy, S.~A. (2003).
\newblock Optimal dynamic treatment regimes.
\newblock {\em J. R. Statist. Soc. B}, 65(2):331--355.

\bibitem[Ning and Liu, 2017]{Ning2017}
Ning, Y. and Liu, H. (2017).
\newblock A general theory of hypothesis tests and confidence regions for
  sparse high dimensional models.
\newblock {\em Ann. Statist.}, 45(1):158--195.

\bibitem[Qian and Murphy, 2011]{qian2011}
Qian, M. and Murphy, S.~A. (2011).
\newblock Performance guarantees for individualized treatment rules.
\newblock {\em Ann. Statist.}, 39(2):1180.

\bibitem[Rosenbaum and Rubin, 1983]{Rosenbaum1983}
Rosenbaum, P.~R. and Rubin, D.~B. (1983).
\newblock The central role of the propensity score in observational studies for
  causal effects.
\newblock {\em Biometrika}, 70:41--55.

\bibitem[Rubin, 1974]{rubin1974}
Rubin, D.~B. (1974).
\newblock Estimating causal effects of treatments in randomized and
  nonrandomized studies.
\newblock {\em J. Educ. Psychol.}, 66(5):688.

\bibitem[Rubin, 2005]{rubin2005}
Rubin, D.~B. (2005).
\newblock Causal inference using potential outcomes.
\newblock {\em J. Am. Stat. Assoc.}, 100(469):322--331.

\bibitem[Shi et~al., 2018]{shi2018}
Shi, C., Fan, A., Song, R., and Lu, W. (2018).
\newblock High-dimensional {$A$}-learning for optimal dynamic treatment
  regimes.
\newblock {\em Ann. Statist.}, 46(3):925--957.

\bibitem[Shi et~al., 2020]{shi2020breaking}
Shi, C., Lu, W., and Song, R. (2020).
\newblock Breaking the curse of nonregularity with subagging---inference of the
  mean outcome under optimal treatment regimes.
\newblock {\em J. Mach. Learn. Res.}, 21(176):1--67.

\bibitem[Shi et~al., 2016]{shi2016}
Shi, C., Song, R., and Lu, W. (2016).
\newblock Robust learning for optimal treatment decision with
  np-dimensionality.
\newblock {\em Electron. J. Statist.}, 10(2):2894--2921.

\bibitem[Song et~al., 2017]{song2017}
Song, R., Luo, S., Zeng, D., Zhang, H.~H., Lu, W., and Li, Z. (2017).
\newblock Semiparametric single-index model for estimating optimal
  individualized treatment strategy.
\newblock {\em Electron. J. Statist.}, 11(1):364--384.

\bibitem[Stone, 1982]{stone1982}
Stone, C.~J. (1982).
\newblock Optimal global rates of convergence for nonparametric regression.
\newblock {\em Ann. Statist.}, 10(4):1040--1053.

\bibitem[Tian et~al., 2014]{Tian2014}
Tian, L., Alizadeh, A.~A., Gentles, A.~J., and Tibshirani, R. (2014).
\newblock A simple method for estimating interactions between a treatment and a
  large number of covariates.
\newblock {\em J. Am. Stat. Assoc.}, 109:1517--1532.

\bibitem[{Van der Vaart}, 1999]{Vaart1999}
{Van der Vaart}, A. (1999).
\newblock Semiparametric statistics.
\newblock In {\em Lectures on Probability Theory and Statistics}, Springer
  series in statistics. Springer-Verlag, Berlin.

\bibitem[Xu et~al., 2015]{Xu2015}
Xu, Y., Yu, M., Zhao, Y., Li, Q., Wang, S., and Shao, J. (2015).
\newblock {Regularized outcome weighted subgroup identification for
  differential treatment effects}.
\newblock {\em Biometrics}, 71(3):645--653.

\bibitem[Yadlowsky et~al., 2020]{Yadlowsky2020}
Yadlowsky, S., Pellegrini, F., Lionetto, F., Braune, S., and Tian, L. (2020).
\newblock Estimation and validation of ratio-based conditional average
  treatment effects using observational data.
\newblock {\em J. Am. Stat. Assoc.}, 0(0):1--18.

\bibitem[Yuan, 2015]{Ming2016}
Yuan, M. (2015).
\newblock Outcome weighted learning with a reject option.
\newblock In Moodie, E. E.~M. and Kosorok, M.~R., editors, {\em Adaptive
  Treatment Strategies in Practice}. Society for Industrial and Applied
  Mathematics, Philadelphia, PA.

\bibitem[Zhang et~al., 2012a]{Zhang2012b}
Zhang, B., Tsiatis, A.~A., Davidian, M., Zhang, M., and Laber, E. (2012a).
\newblock {Estimating optimal treatment regimes from a classification
  perspective}.
\newblock {\em Stat}, 1(1):103--114.

\bibitem[Zhang et~al., 2012b]{zhang2012robust}
Zhang, B., Tsiatis, A.~A., Laber, E.~B., and Davidian, M. (2012b).
\newblock A robust method for estimating optimal treatment regimes.
\newblock {\em Biometrics}, 68(4):1010--1018.

\bibitem[Zhao et~al., 2019]{zhao2019}
Zhao, Y., Laber, E.~B., Ning, Y., Saha, S., and Sands, B.~E. (2019).
\newblock Efficient augmentation and relaxation learning for individualized
  treatment rules using observational data.
\newblock {\em J. Mach. Learn. Res.}, 20(48):1--23.

\bibitem[Zhao et~al., 2012]{Zhao2012}
Zhao, Y., Zeng, D., Rush, A.~J., and Kosorok, M.~R. (2012).
\newblock {Estimating Individualized Treatment Rules Using Outcome Weighted
  Learning}.
\newblock {\em J. Am. Stat. Assoc.}, 107(499):1106--1118.

\end{thebibliography}
\end{document}